\documentclass[journal]{IEEEtran}
\usepackage{blindtext}
\usepackage{graphicx}
\usepackage{xcolor}
\usepackage{scrextend}
\usepackage{amsmath}
\usepackage{mathtools, cuted}
\usepackage{lipsum, color}
\usepackage{url}
\usepackage{mathtools,amssymb,mathtools,mathdots,bm,fixltx2e}

\usepackage{graphicx} 
\usepackage{algorithmicx}
\usepackage[noend]{algpseudocode}

\usepackage{amsmath}
\usepackage{algorithm}
\usepackage[noend]{algpseudocode}

\makeatletter
\def\BState{\State\hskip-\ALG@thistlm}
\makeatother
\addtokomafont{labelinglabel}{\sffamily}

\graphicspath{ {figs/} }

\usepackage{multirow}
\usepackage{xcolor,colortbl}

\usepackage{caption}
\usepackage{subcaption}
\usepackage{mwe}
\usepackage{fixltx2e}

\definecolor{Gray}{gray}{0.85}
\definecolor{LightCyan}{rgb}{0.8,1,1}

\newcolumntype{g}{>{\columncolor{Gray}}c}
\newcolumntype{w}{>{\columncolor{white}}c}
\newcolumntype{b}{>{\columncolor{LightCyan}}c}

\ifCLASSINFOpdf
\else
\fi

\begin{document}
%

\title{Vehicular Communication and Mobility Sustainability: the Mutual Impacts in Large-scale Smart Cities}
%
%
%

\author{Ahmed~Elbery, Hesham~Rakha, \textit{Member, IEEE}, and Mustafa~ElNainay, \textit{Senior  Member, IEEE}
}

\maketitle

\begin{abstract}

Intelligent Transportation Systems (ITSs)  is the backbone of transportation services in smart cities. ITSs produce better-informed decisions using real-time data gathered from connected vehicles. In ITSs, Vehicular Ad hoc Network (VANET) is a communication infrastructure responsible for exchanging data between vehicles and Traffic Management Centers (TMC). VANET performance  (packet delay and drop rate) can affect the performance of ITS applications. Furthermore, the distribution of communicating vehicles affects the VANET performance. So, capturing this mutual impact between communication and transportation is crucial to understanding the behavior of ITS applications.
Thus, this paper focuses on studying the mutual impact of VANET communication and mobility in city-level ITSs.  We first introduce a new scalable and computationally fast framework for modeling large-scale  ITSs including communication and mobility. In the proposed framework, we develop and validate a new mathematical model for the IEEE 802.11p MAC protocol which can capture the behavior of medium access and queuing process.  This MAC model is then integrated within a microscopic traffic simulator to accurately simulate vehicle mobility. This integrated framework can accurately capture the mobility, communication, and the spatiotemporal impacts in large-scale ITSs. Secondly, the proposed framework is used to study the impact of communication on eco-routing navigation performance in a real large-scale network with real calibrated vehicular traffic.
The paper demonstrates that communication performance can significantly degrade the performance of the dynamic eco-routing navigation when the traffic density is high. It also shows that the fuel consumption can be increased due to lack of communication reliability.


\end{abstract}


%
\IEEEpeerreviewmaketitle

\section{Introduction}

\IEEEPARstart{I}{}ntelligent Transportation Systems (ITSs) are expected to be the core of future transportation systems. In an ITS, networked sensors, microchips, and communication devices work together to collect, process, and disseminate information about the transportation system. Consequently, enabling the traffic management center (TMC) to make better-informed decisions to improve the performance of the overall transportation system. These decisions can be made to improve traveler transportation mode selection, departure time selection, route selection, or traffic signal timing optimization. The correctness and the accuracy of these decisions depend on the accuracy of the data collected at the TMC. The communication network is a major component in such feedback ITS applications. It is responsible for exchanging data between the different sensors/actuators in the network and the TMC. 

Vehicular ad hoc network (VANET) \cite{hafeez2013performance} is a promising communication technology that is expected to serve as communication backbone of ITS systems. Consequently, when studying an ITS application performance, it is imperative to study the impact of VANET communication reliability parameters (packet delay and drop rate) on mobility sustainability and performance of ITS.

Moreover, communication system performance is affected by mobility parameters (speed, traffic demand rate, and congestion) in a road network. 
This bidirectional interdependence between mobility and communication creates a mutual impact loop between them, where mobility affects communication and communication affects mobility. 

Dynamic eco-routing navigation is a good example for this mutual loop. 
Eco-routing attempts to route vehicles along fuel-efficient routes  utilizing fuel consumption cost information collected in real-time from vehicles currently moving on the network. Using these real-time link costs, the TMC  assigns vehicles the routes that minimizes their fuel consumption. 

Eco-routing navigation includes both directions of influence where vehicles' mobility and distribution affect the communication performance, and on the other direction, the communication performance  can influence the route assignment, consequently, the mobility and distribution of the vehicles.

However, studying and modeling such systems are challenging not only because of the intricate interdependency of the communication and the mobility components, but also because of the scale at which the ITS works. Most of the ITS applications operate on a city level road network that mandates the modeling of tens or hundreds of thousands of concurrent vehicles, which makes such studies more challenging.

The main objective of this paper is to study the mutual impact of VANET communication reliability and the mobility sustainability in city-level ITSs, where eco-routing is used as an ITS applications.  
 
To achieve this objective, we  propose a new scalable framework that is capable of modeling the mutual interaction of the communication and transportation systems in the vehicular environment. The proposed model is characterized by its scalability that supports tens of thousands of concurrent vehicles in the network. It is also capable  of estimating the packet delay and drop probability based on the number of vehicles in the communication area and background packet generation rate.

To build this framework, we firstly develop a model for the Medium Access Control (MAC) technique with a finite buffer size in vehicular networks based on the IEEE802.11p standard . This model utilizes both the Markov chain and queuing theory to estimate the packet delay and drop probability. Secondly, this model is validated against a benchmark simulation data generated using OPNET simulator. Thirdly, to allow for studying the mutual impact of the communication and transportation systems, this model is incorporated in a transportation microscopic traffic simulation software; INTEGRATION \cite{rakha2012integration}. 

Then, the developed simulation framework is used to quantify the impact of VANET communication on the eco-routing navigation performance in a real large-scale network; the downtown area in the city of Los Angles (LA) which is approximately 10 km x 15 km. The Road network is built using the GIS shapefiles. The car traffic demand is calibrated to existing conditions using data obtained from different sources. 
To the best of our knowledge, this is the first  attempt at  considering and modeling the mutual interaction of the communication and transportation systems in such real large networks with real traffic. This is also the first attempt at quantifying the impact of the communication system on dynamic feedback eco-routing system.

This paper continues by providing an overview of VANET communication and introducing a novel communication model for vehicular communication based on the IEEE 802.11p standard \cite{IEEE_P1609, IEEE_P1609_2015}. After validating the model, the paper describes the eco-routing application and how the communication model was incorporated in the INTEGRATION software \cite{rakha2012integration}, which is a microscopic traffic and route assignment simulator. Then, the simulation network and the simulation results are presented and discussed followed by the study conclusions.

 \section{VANET Communication}
\begin{figure}[b]
    \includegraphics[scale=0.3]{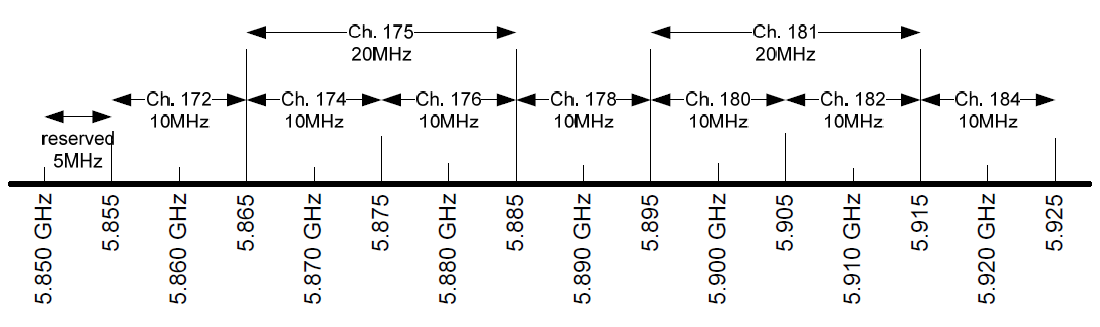}
    \centering
    \caption{DSRC channels in the U.S.}
    \label{fig:VANET_Channels}
\end{figure}

Using VANET, moving vehicles can communicate with other vehicles (vehicle-to-vehicle [V2V]), roadside units (RSUs) (vehicle-to-infrastructure [V2I]), or even hand-held mobile devices (vehicle-to-device [V2D]) using a Dedicated Short Range Communication (DSRC) system. DSRC has been standardized by IEEE 802.11p in the 1609 family of standards known as Wireless Access in Vehicular Environments (WAVE) \cite{IEEEStdWAVE}. The DSRC spectrum is 75 MHz assigned by the U.S. Federal Communication Commission to the vehicle-to-X (V2X) communications. Counterpart agencies in other countries have made similar allocations. This spectrum is divided into seven 10-MHz wide channels. The frequency band 5.850 GHz to 5.855 GHz is reserved as shown in Fig \ref{fig:VANET_Channels}. The middle channel, channel 178 shown in Fig \ref{fig:VANET_Channels},  is the control channel (CCH). Channels 172, 174, 176, 180, 182, and 184 are service channels (SCH) that are intended for general purpose application data transfers. In our application, eco-routing can use any of the general purpose service channels.

While the physical layer of IEEE 802.11p is similar to the IEEE 802.11a standard \cite{IEEE802.11a, IEEE802.11p-2010}, IEEE 802.11p uses an Enhanced Distributed Channel Access (EDCA) media access control (MAC) sub-layer protocol that is based on the IEEE 802.11e standard \cite{IEEE802.11e} with some modifications to the transmission parameters. IEEE 802.11p supports transmission rates of up to 54 Mb/s within ranges up to 1,000 m. In IEEE 802.11p, the EDCA provides quality-of-service (QoS) by supporting four different traffic Access Categories (ACs), namely, Background (BK), Best-effort (BE), Video (VI), and Voice (VO). Each AC has different medium access parameters such as Arbitration Inter Frame Space (AIFS) and Contention Window (CW) limits. In this paper, to enable large-scale modeling of both communication and transportation systems, we simplify the model by assuming only one AC. This assumption will be discussed in more detail and will be validated in Section \ref{assumption}.





\subsection{WAVE and Eco-routing}

In this paper, we focus on a dynamic feedback-based eco-routing application, which is described in detail in Section \ref{sec:eco}. This subsection describes the eco-routing application network requirements. 

The eco-routing application sends only one packet for each road link whose traversal time ranges between few seconds to a number of minutes. Consequently, use of TCP would be wasting resources. Regarding the reliability requirements, our previous work \cite{elbery2015eco} on a small network showed that the eco-routing application is robust to packet drops and delays, the results in \cite{elbery2015eco} show that a drop rate up to  25\% does not have a significant impact on the eco-routing system performance as long as the network is almost covered.  Based on these conclusions, the eco-routing application can safely use the UDP protocol. So, we do not need to consider the TCP protocol in this paper.

Routing is a main functionality of the Internet protocol (IP) \cite{rfcIP,IPv6rfc} that should be considered when studying any network application. However, routing in VANET is still an open and wide area of research and the performance of the different routing techniques are still under investigation by the research community. Thus, in this paper we do not consider the routing protocols and their impact on the eco-routing system performance. Consequently, we assume only a V2I communication. However, our proposed model can be easily modified to include the impact of multi-hop communication in the case of routing. The proposed model can also consider the packet traffic load introduced by the routing protocol as an input parameter to our model.

The WAVE MAC protocol has a direct mutual interaction with the transportation network where it can be significantly affected by the transportation network parameters. For example, in the case of high traffic congestion, the large number of stations (vehicles) attempting to send data can result in higher competition in the wireless medium, resulting in higher probabilities of packet collisions,  higher drop probabilities, and longer packet delivery delays. Moreover, the MAC protocol is also affected by the communication parameters, for instance the application packet generation rate (accompanied with the number of station) has significant impact on the delay and drop rate in the MAC protocol. On the other hand,  the drop rate and packet delay in the MAC protocol can influence the performance of the eco-routing application in the ITS.  Consequently, the MAC protocol is a good representative of the mutual impact of communication and transportation systems.   So, in this paper, we focus on the MAC protocol in WAVE and its  impact on the eco-routing application. The next subsection briefly describes the MAC protocol in vehicular environment.

\subsection{WAVE Medium Access Technique }

According to the IEEE 802.11p standard, within a station, every AC acts as a stand-alone virtual station that has its own queue and its own access parameters. Whenever any AC has a frame to send, it initializes its back-off counter to a random value within the range of the initial $CW$  ($w_0$). Then, when it detects that the medium is idle for a specific time duration (usually equal to the AIFS), it counts down the back-off counter. If the medium is busy, the counter will be held. The station can send the frame only when the back-off counter is zero. When more than one EDCA count their back-off timers to zero and attempt to transmit at the same time, a virtual collision takes place within the same station, which is referred to as internal collision. In this case, access to the medium will be granted to the highest priority AC, while the lower priority colliding AC doubles its CW and backs off again. 

If two stations start transmitting in the same time slot, an external collision will take place and both transmitted signals will be destroyed. Because the sender cannot detect the collision while transmitting, it has to wait for the acknowledgment (ACK). If an ACK is not received within a specified time, the sender assumes a collision and doubles its CW and backs off again. This process can be repeated until reaching a retransmission attempt threshold \( (M+f) \). During these attempts, the CW can be increased up to the maximum $CW$ ($w_{max}$). Consequently, the number of times the CW can be increased is \(M=log_2 (w_{max}/w_0 )\), and $f$ is the number of retransmission attempts allowed after reaching ($w_{max}$).

\section {Proposed Medium Access Model}
This section describes the analytical model we developed to represent the single AC operation of the IEEE802.11p. 

Compared to the previous models in the literature, our proposed model is characterized by combining the following features:
\begin{itemize}
	\item It considers the MAC layer queue size.
    \item It is capable of estimating both the average processing and queuing delay for the packet.
	\item It works for both saturated and unsaturated cases.
	\item  It limits the number of retransmission attempts. When a packet reaches this limit, it will be dropped.
\end{itemize}

First, our proposed model considers the MAC layer queue size and its impact on communication performance. Most of the previous models that were developed for both IEEE 802.11a, IEEE 802.11e, and IEEE 802.11p do not consider the queue size, the queuing process, and their impact on the performance of IEEE 802.11. For example, Bianchi\( ^ {^,} \)s model \cite{bianchi2000}, which was proposed in 2000 and then refined in 2010 \cite{tinnirello2010}, as well as models proposed in \cite{chatzimisios2004, eichler2007, chen2011,han2012, zheng2016performance, weng2016,hajlaoui2016} do not consider queuing. 
Some models consider queuing but assume an infinite queue size such as Engelstad et al. \cite{engelstad2005}, which modeled the EDCA in IEEE 802.11e with an infinite queue, meaning the packet always finds a buffer to be stored in, which is not a realistic assumption. Moreover, the queue size can have a significant impact on the performance of IEEE 802.11 communication. For instance, the smaller the queue size, the lower the number of packets can be queued. In the case of high packet traffic rates, many of the packets will be rejected by the queue, which will increase the packet drop ratio. On the other hand, larger queue size will result in increasing the queuing delay to very high values. 
In contrast to these models, our model assumes a finite queue size of length $K$ in the MAC layer, which makes it more realistic. To model the queuing process, in the proposed model, we use the M/M/1/K queuing model \cite{kleinrock1976}. This queuing model is incorporated with the MAC protocol so that the back-off technique and the queue interact with each other. Consequently, with the help of this queuing model, we were able to compute both the queuing and processing delay. In addition, the queuing model parameters were used to estimate the throughput and the packet drop rate.

Secondly, compared to most of the previous research, our proposed model supports both a saturated and unsaturated wireless data network. For example, the models in \cite{bianchi2000,tinnirello2010,chatzimisios2004,eichler2007,chen2011,han2012,weng2016,hajlaoui2016} assume saturated conditions in the modeling of the DCA/ECDA using a Markov chain. Alternatively, a model developed by Engelstad et al. \cite{engelstad2005} for both unsaturated and saturated traffic conditions, assumes an infinite queue so it cannot realistically estimate the delay and throughput.

\begin{figure*}[h!]
	\includegraphics[height=0.7\textwidth]{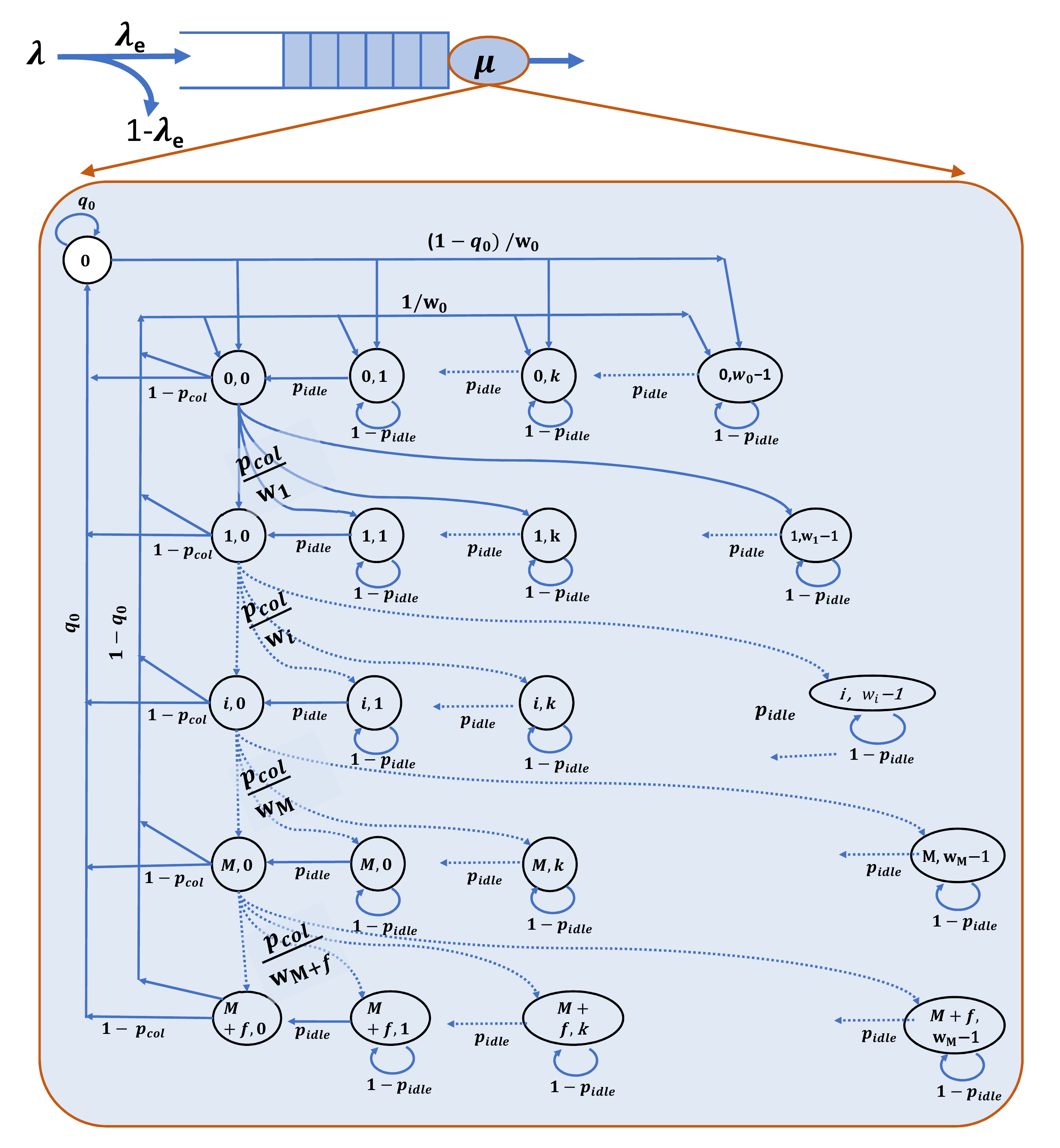}
	\centering
	\caption{Markov chain model for the medium access.}
	\label{fig:DTMC}
\end{figure*}

Thirdly, many of the models in the literature do not limit the number of retransmissions for each individual packet. This is basically to simplify the mathematical derivation of the model. Our proposed model assumes a limited number of retransmission attempts, after which the packet is dropped. These three characteristics are illustrated in Fig. \ref{fig:DTMC} and the model is described in Subsection \ref{subsec:Derivation}.

\subsection{Model Assumptions}
\label{assumption}
The main purpose of this paper is to enable large-scale modeling of the communication system in a vehicular environment. Our previous research in \cite{elbery2015eco,elbery2015}  shows that the discrete event simulations of communication in large-scale networks is computationally expensive in terms of both memory and simulation time increasing exponentially with the number of vehicles on the network \cite{elbery2015}.  In this paper, instead of using a discrete event simulation for the communication, we develop a communication model that can estimate the average packet drop probability and the average total delay.

\begin{figure}[h!]
    \includegraphics[scale=0.8]{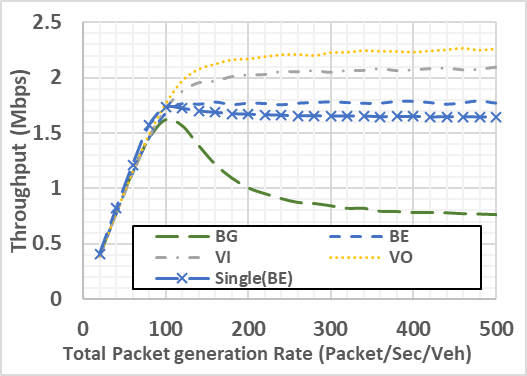}
     \centering
     \caption{Comparison between the BE traffic using single AC and 4 ACs.}
     \label{fig:assum_jsut}
\end{figure}

To allow for such large-scale systems, we simplified this model to assume a single AC instead of the four ACs in VANET. This assumption is based on a comparative simulation study between the single AC and multiple ACs in VANET.  We ran simulations using the OPNET simulator to compare the network performance in case of a single AC and four ACs for different traffic rates. In the case of four ACs, each AC generates the same traffic rate $\lambda$. In the case of single AC, a traffic rate of $4\lambda$ is generated and assigned to the default best-effort AC. Then, the throughput of the single AC is divided by $4$ and compared to the data traffic throughput in case of the four ACs. Fig. \ref{fig:assum_jsut} shows the comparison of the data traffic in the case of four ACs and $\frac{1}{4}$ the throughput of the single AC. Fig. \ref{fig:assum_jsut} shows that the throughput of the BE AC is very close to the approximated BE using a single AC. The comparison shows that the error is less than 11\%. Based on this analysis and the results in Fig. \ref{fig:assum_jsut}, we can use the single AC to represent the BE AC in the full fledged model. 



\subsection{Model Derivation}
\label{subsec:Derivation}

In building a model for the MAC layer, a two-dimensional Markov chain, shown in Fig.  \ref{fig:DTMC}, is utilized.  The state 0 in the model represents the system-empty state when both the system and the queues are empty. Each of the other states is defined by \((i,j)\), where $i$ and $j$ are the back-off stages and back-off counter value, respectively. Table \ref{table:model_parameters} shows the symbols used in this model.

\begin{table}
\centering
\caption{The model parameters}
\label{table:model_parameters}
\begin{tabular}{| l| p{70mm}|} 
\hline
\textit{\textbf{Symbol}}     & \textit{\textbf{Description}}                                                \\ \hline
\(i  \)                      &The back-off stage number        \\ \hline
\(j  \)                      & The back-off counter                \\ \hline
\(M  \)                      & The maximum number of times of increasing the CW     \\ \hline
\(f  \)                      & The maximum number of times of retransmission without increasing the CW     \\ \hline
\(w_i\)                      & The CW range for stage \(i\)           \\ \hline
\(w_0\)                      & The initial value for the maximum CW \(i\)           \\ \hline
\(\alpha\)                   & The CW increasing factor, where \( w_i= w_0  \alpha^i\). Its typical value is 2.              \\ \hline
\(p_{idle_{slot}}  \)        & The probability that a medium is idle in any time slot        \\ \hline
\(p_{idle} \)                & The probability that the medium is idle                                      \\ \hline
\(q_0\)                      & The probability that the system is empty (no packet in the system)            \\ \hline
\(p_{suc}  \)                & The probability that a medium is occupied with a successful transmission             \\ \hline
\(p_{fail}  \)               & The probability that a medium is occupied with a failure transmission              \\ \hline
\(p_{tran}  \)               &The probability that a station starts transmission in any time slot          \\ \hline
\(p_{col}  \)                & The probability that the packet collides                                     \\ \hline
\(P(i,j)  \)                 & The probability that the system is in state \( (i,j) \)      \\ \hline
\(\lambda  \)                & The packet arrival rate       \\ \hline
\(\mu  \)                    & The packet service rate      \\ \hline
\(T_{serv}  \)               & The packet service time     \\ \hline
\(T_q  \)                    & The packet queuing time    \\ \hline
\(N  \)                      & The number of vehicles (stations) in the communication range   \\ \hline
\(T_{slot}  \)               & The length of the time slot (sec)       \\ \hline
\(T_s  \)                    & The transmission time of a successful frame transmission     \\ \hline
\(T_f \)                     & The transmission time of a failure frame  transmission         \\ \hline
\(AIFS  \)                   & The number of time slots for the Arbitration Inter Frame Space       \\ \hline
\(\rho  \)                   & The traffic intensity for the queuing model                                \\ \hline
\(K  \)                      & The size of the buffer                                                       \\ \hline
\(Q_n\)                      & The probability that the queue has $n$ packets                       \\ \hline
\( \lambda_{eff}\)           & The effective packet generation rate                       \\ \hline

\end{tabular}
\end{table}

To solve this model we derive all the state probabilities $P(i,j)$ in addition to $P(0)$ as a function of $P(0,0)$. The summation of all these state probabilities should equal 1.0.

From the Markov chain in Fig. \ref{fig:DTMC}, we can find that the probability that the system is in state 0 as

\begin{equation}
  \label{eq:P(0)1}
  \begin{multlined}
     P(0)=   \frac{q_0}{1-q_0}   \Big( P(M+f-1,0)\\   +  (1-p_{col} ) \sum_{i=0}^{M+f-2} P(i,0) \Big ).
  \end{multlined}
\end{equation}

And \(P(0,j)\) can be expressed as
\begin{equation}
      \label{P(0,j)}
	\begin{multlined}
		P(0,j)= \frac{w_{0}-j}{w_0} \; \frac{1-q_0}{p_{idle}}\Bigg(  P(0)+ P(M+ f-1,0)+ \\  (1-p_{col} )\sum_{i=0}^{M+f-2}{P(i,0)}  \Bigg ) \quad  , j= 1,2, ...  ,w_{0} -1
      \end{multlined}
\end{equation}

and \(P(0,0)\) is computed as
\begin{equation}
	\label{eq:P(0,0)}
	\begin{multlined}
	       P(0,0)= (1-q_0)   \Bigg ( P(0)+ P(M+ f-1,0)\\+  (1-p_{col} )\sum_{i=0}^{M+f-2}{P(i,0)}  \Bigg ).
	\end{multlined}
\end{equation}

From Equations \ref{P(0,j)} and \ref{eq:P(0,0)}, we can compute \(P(0,j)\) as
\begin{equation}
	\label{eq:P(0,j)}
	P(0,j)=\frac{w_0-j}{w_0}  \frac{1}{p_{idle}} P(0,0)     \;\; \quad   , j=1, 2, ..., w_0-1.
\end{equation}

Consequently,  \(P(i,0)\) can be calculated as
\begin{equation}
\label{eq:P(i,0)}
P(i,0)=p_{col}^{i} \;   P(0,0)   \quad    ,   i=0, 1, .... , M+f-1
\end{equation}

and

\begin{equation}
\label{eq:P(i,j)}
	\begin{multlined}
		P(i,j)=  \frac{w_i-j}{w_i}   \frac{p_{col}^i}{p_{idle  }} P(0,0)  \\ , i=1,2, ... ,M+f-1 \quad and \quad   j=1, 2, ..., w_i-1.
	\end{multlined}
\end{equation}

From Equations \ref{eq:P(0)1} and \ref{eq:P(i,0)} we can find the relation between \(P(0)\)  and \(P(0,0)\) as
\begin{equation}
\label{eq:P(0)2}
P(0)=  \frac{q_0}{1-q_0}  P(0,0).
\end{equation}


Now we have all the state probabilities expressed in terms  of \(P(0,0)\). Since the summation of all the probabilities equals 1, then

\begin{equation}
\label{eq:sum}
P(0)+ \sum_{i=0}^{M+f-1}P(i,0) +  \sum_{k=1}^{w_0-1}P(0,k)   + \sum_{i=1}^{M+f-1} \sum_{k=1}^{w_{i-1}}P(i,k) =1.
\end{equation}

Notice that the window exponential factor is \(\alpha \) for \( i \leq M \), i.e  

\begin{equation}
\label{eq:w_i}
 w_i= \left\{\begin{matrix}
 w_0 \; \alpha^ i &  &  & i \leq  M ;  \\
                  &  &  &            \\
 w_0 \; \alpha^ M &  &  & i  >  M.
\end{matrix}\right.
\end{equation}

From Equations \ref{eq:sum} and \ref{eq:w_i}, we can calculate \(P(0,0)\) as shown in Equation \ref{eq:P(0,0)2}. 

Now, to solve this model, we have to calculate the values of \(p_{col}, p_{idle}\), and \(q_0\). To do that, we need to find a relationship between these three parameters and the state probabilities.

	\begin{equation}
	\label{eq:P(0,0)2}
	\begin{multlined}
	P(0,0) =\Bigg( \frac{q_0}{1-q_0} +    \frac{1-{p_{col}}^{M+f}}{1-  p_{col}}+  \frac{w_0-1}{2 \;  p_{idle}} +   \\  \frac{1}{2 \; p_{idle}}  \Big[ (\alpha ^{M-1} \; w_0 -1   )  \frac{{p_{col}}^{M-1} -{p_{col}}^{M+f} }{1-p_{col}} \\ +	w_0 \frac{\alpha  \; p_{col} - (\alpha  \;  p_{col})^{M-1} }{1- \alpha  \; p_{col}}    +   \frac{ p_{col} - ( p_{col})^{M-1} }{1-  p_{col}}\Big] \Bigg)^{-1}.
	\end{multlined}
	\end{equation}

A collision will occur when two or more stations start transmission in the same time slot. Let the probability that a station starts transmitting at a time slot be \(p_{trans}\), then
\begin{equation}
    \label{eq:ptran}
    p_{trans}=  \sum_{i=0}^{M+f-1}P(i,0).
\end{equation}

When a station sends a packet, the probability that this packet collides is
\begin{equation}
    \label{eq:pcol}
    p_{col}= 1-(1-p_{tran} )^{N-1}.
\end{equation}
So, for the entire system, the medium will be idle at any time slot only when no station is sending 
\begin{equation}
    \label{eq:pidleslot}
    p_{idle_{slot}}=  (1-p_{tran} )^{N}.
\end{equation}

The station decides that the medium is idle (\(p_{idle }\))  after \(AIFS\) idle time slots in raw

\begin{equation}
    \label{eq:pidle}
    p_{idle }= {p_{idle_{slot}}}^{AIFS}.
\end{equation}

For the entire system, the probability that a packet is successfully transmitted without collision is computed as

\begin{equation*}
     			p_{suc}=  \binom{N}{1} \; p_{tran  } \; (1-p_{col})  = N \; p_{tran}\; (1-p_{col} )
\end{equation*}

\begin{equation}
     \label{eq:psuc}
     		   p_{suc}=   N\;  p_{tran} \;  (1-p_{tran} )^{N-1}.
\end{equation}

Using Equations \ref{eq:ptran} through \ref{eq:psuc}, the two parameters \(p_{col} \) and \(p_{idle}\) can be computed in terms of state probabilities and therefore in terms of \(P(0,0)\). The only missing part is finding a relation between \(q_0\) and the state probabilities. To solve for \(q_0\), we use the \(M/M/1/K\) model, where we assume the packet inter-arrival time between packets is totally random (i.e. exponentially distributed) with an average rate \(\lambda\). Assuming that the service rate is \(\mu=\frac{1}{T_{serv}}\), the service time \(T_{serv}\)  is the average time elapsed in processing the packet. It is the summation of the average time the packet stays in each stage. The packet can stay a time \(T_w\) in every state plus the average frame transmission time \(T_{tr_{av}}\), which can be calculated as

\begin{equation}
     \label{eq:T_w} 
     T_w= p_{fail} \; T_f  + p_{suc} \; T_s + \frac{1}{p_{idle}} T_{slot}.
\end{equation}

\begin{equation}
     \label{eq:T_tr_av}
    T_{tr_{av}}=p_{col} \;  T_f + (1-p_{col} ) \; T_s.
\end{equation}
where
 $p_{fail}= 1-p_{suc}-p_{idle_{slot}} $.
  \(T_s\) and  \(T_f\) are the successful transmission time and the failure transmission time respectively.  \(T_s\)  and  \(T_f\) depends on whether MAC uses the basic or  advanced (RTS/CTS) access modes.
  
In case of the basic access mode, the transmitting node directly sends its packet
\begin{equation}
     \label{eq:T_s} 
    T_s=T_{AIFS}+T_{frame}+T_{pro}+ T_{SIFS}+T_{ack}+ T_{pro},
\end{equation}

\begin{equation}
     \label{eq:T_s2}
    T_f=T_{AIFS}+T_{frame}+T_{pro}.
\end{equation}

In the case of the RTS/CTS access mode, the transmitting node firstly sends a request to send packet and waits for the clear to send packet, after the successful handshaking, the sender can send its data frame. In this case, the successful and failed transmission times are calculated as:
\begin{equation}
     \label{eq:T_s3}
	\begin{multlined}
      T_s=T_{AIFS}+  T_{rts}+ T_{pro}+ T_{SIFS}+T_{cts}+T_{pro} \\+ T_{SIFS}+ T_{frame}+T_{pro}+ T_{SIFS}+T_{ack}+ T_{pro},
     \end{multlined}
\end{equation}

\begin{equation}
     \label{eq:T_s4} 
      T_f=T_{AIFS}+  T_{rts}+ T_{pro}.
\end{equation}

The term \(T_w\) is the time required by the station while sensing the medium to ensure it is idle. Consequently, the service time \(T_{serv}\) can be calculated as
\begin{equation}
     \label{eq:Tserv} 
      T_{serv}= T_{slot} \sum_{i=0}^{M+f-1} \left (   \frac{T_w (w_i-1)}{2}+T_{tr_{av}}  \right ){p_{col}}^i.
\end{equation}

Now we can calculate the traffic intensity \(\rho=\frac{\lambda}{\mu}\), and subsequently \(q_0\) as

\begin{equation}
     \label{eq:q_0} 
      q_0=\left\{\begin{matrix}
 		\frac{1-\rho}{1-\rho^{K+1}}  &  \lambda\neq  \mu ;  \\
           &  \\
 		\frac{1  }{K+1}  &  \lambda =   \mu. 
\end{matrix}\right.
\end{equation}

Now we can solve this model and estimate the total communication throughput and delay. The total network throughput \(Thr\) can be calculated as
\begin{equation}
     \label{eq:Thr} 
     Thr =N(1-q_0)\big(1-P(M+f-1,0)  p_{col} \big)(1-p_{fail} ).
\end{equation}

\(Thr\) is simply the multiplication of a set of terms: \((1-q_0)\) represents the time ratio at which the system has at least a packet, \(\big(1-P(M+f-1,0)  p_{col} \big)\) describes the probability that this packet was not dropped due to the maximum retransmission attempts constraint, and \((1-p_{fail} )\) is the portion of time that is not occupied by collisions.

\subsection{Communication Model Validation}
\begin{figure*}
   \centering
    \begin{subfigure}[t]{0.235\textwidth}
    	\centering
\includegraphics[width=4.5cm,height=3.5cm]{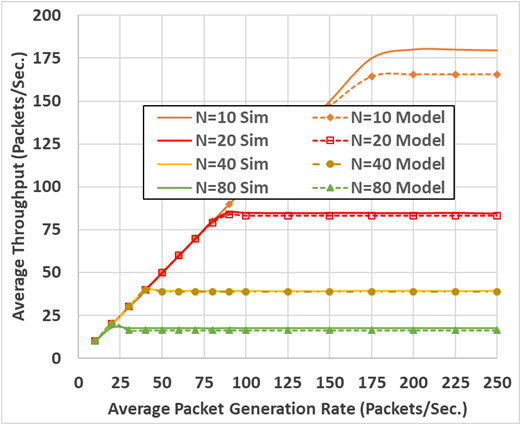}
        \caption{Basic Access Mode, Packet Size = 500 Bytes}
    \end{subfigure}%
    ~
    \begin{subfigure}[t]{0.235\textwidth}
    	\centering
    	\includegraphics[width=4.5cm,height=3.5cm]{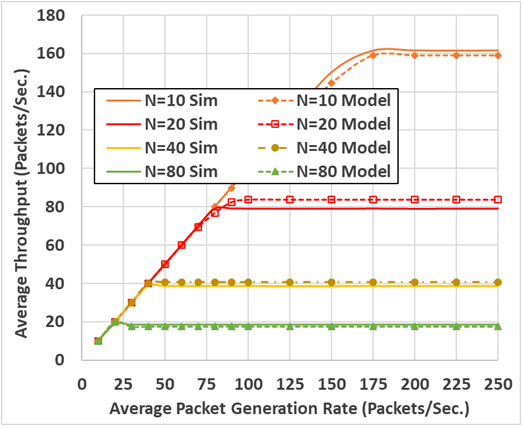}
        \caption{RTS/CTS Access Mode, Packet Size = 500 Bytes}
    \end{subfigure}
    ~
    \begin{subfigure}[t]{0.235\textwidth}
    	\centering
    	\includegraphics[width=4.5cm,height=3.5cm]{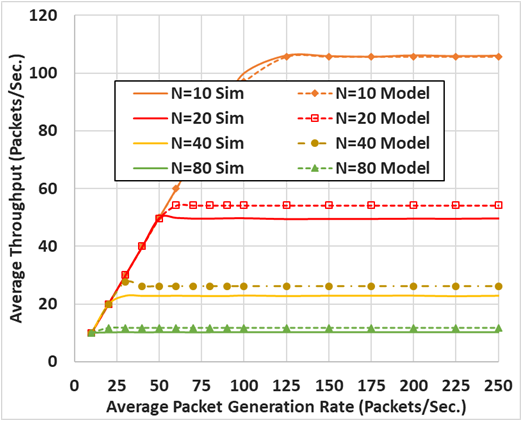}
        \caption{Basic Access Mode, Packet Size = 1000 Bytes}
    \end{subfigure}%
    ~
    \begin{subfigure}[t]{0.235\textwidth}
    	\centering
    	\includegraphics[width=4.5cm,height=3.5cm]{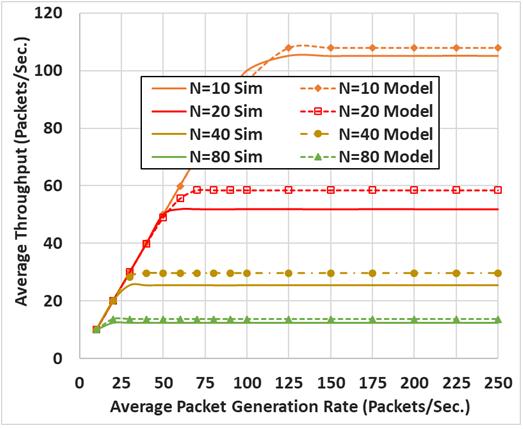}
        \caption{RTS/CTS Access Mode, Packet Size = 1000 Bytes}
    \end{subfigure}
    \caption{Average throughput per vehicle (Packets/Second) versus packet generation rate (Packets/Second), comparing the model to the simulation for different number of vehicles}
	\label{fig:thr}
\end{figure*}

\begin{figure*}
    \centering
    \begin{subfigure}[t]{0.235\textwidth}
        \centering
        \includegraphics[width=4.5cm,height=3.5cm]{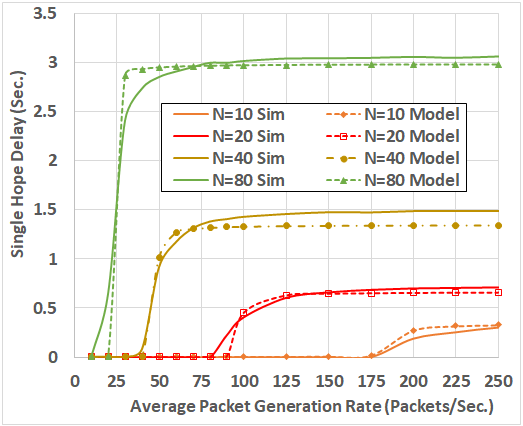}
        \caption{Basic Access Mode, Packet Size = 500 Bytes}
    \end{subfigure}%
    ~
    \begin{subfigure}[t]{0.235\textwidth}
        \centering
        \includegraphics[width=4.5cm,height=3.5cm]{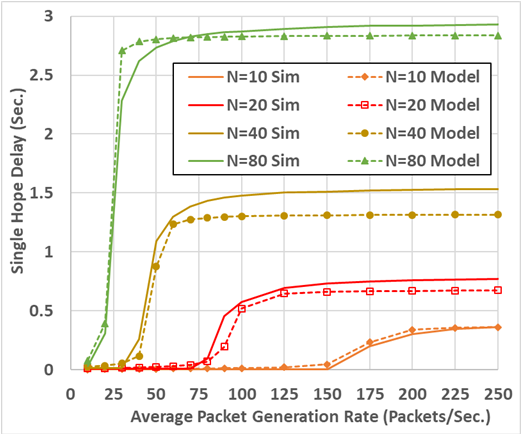}
        \caption{RTS/CTS Access Mode, Packet Size = 500 Bytes}
    \end{subfigure}
    ~
    \begin{subfigure}[t]{0.235\textwidth}
        \centering
        \includegraphics[width=4.5cm,height=3.5cm]{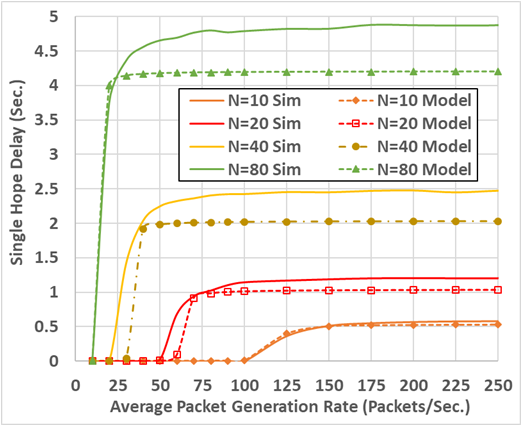}
        \caption{Basic Access Mode, Packet Size = 1000 Bytes}
    \end{subfigure}%
    ~
    \begin{subfigure}[t]{0.235\textwidth}
        \centering
        \includegraphics[width=4.5cm,height=3.5cm]{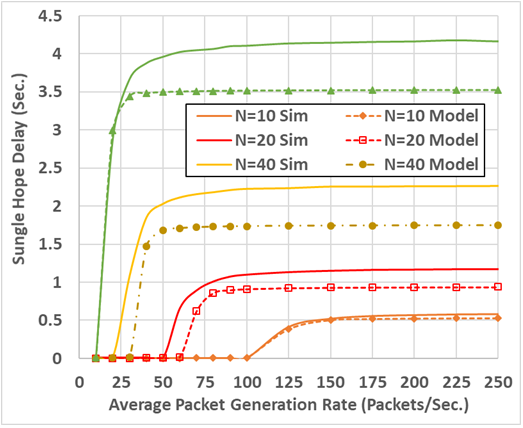}
        \caption{RTS/CTS Access Mode, Packet Size = 1000 Bytes}
    \end{subfigure}
    \caption{Average single hop delay, model versus simulation}
    \label{fig:delay}
\end{figure*}

To validate this communication model, we ran extensive simulations for different vehicular traffic demand levels and number of communication stations considering V2I communication.  We used the OPNET software, which is known currently as the Riverbed modeler \cite{OPNET}. The OPNET modeler is a powerful discrete event simulation tool for specification, simulation, and analysis of data and communication networks. The most important OPNET characteristic is that its results are trusted because its implementations of the standard protocols are tested and validated before publishing. Current versions of OPNET support WAVE as an extension of the IEEE 802.11 implementation. For each simulation scenario we calculated the average network throughput and the average packet delay. The results show an accurate estimation of both throughput and delay of our model compared to the OPNET simulated results (SIM), as shown in Fig. \ref{fig:thr} and Fig. \ref{fig:delay}.

\section{Transportation Traffic Modeling}
This section describes how the developed communication model is integrated into the transportation simulator to capture the inter-dependency of communication and mobility. It first gives and an overview of the INTEGRATION software followed by a description of the eco-routing application. Finally, this section details the implementation and the operation of eco-routing application with communication modeling.

The transportation network is the environment where the vehicular communication takes place. Thus, it is essential to integrate the proposed communication model into a scalable transportation simulation software. After validating the communication model, we implemented and incorporated it within the INTEGRATION software \cite{rakha2012integration}. INTEGRATION is a trip-based microscopic traffic assignment, simulation, and optimization model. INTEGRATION is capable of modeling networks with hundreds of thousands of cars. It is characterized by its accuracy, which comes from its microscopic nature, and its small time granularity; by tracking individual vehicle movements from a vehicle's origin to its final destination at a level of resolution of 1 deci-second. Using this time resolution, INTEGRATION can accurately track vehicles by modeling vehicle car-following, lane-changing, and gap acceptance behavior. The INTEGRATION simulation model provides 10 basic user equilibrium traffic assignment/routing options \cite{rakha2012integration}. Recently, a system equilibrium routing model was added \cite{elbery2017novel}. One important feature of INTEGRATION is its support for eco-routing traffic assignment, which tries to minimize the fuel consumption and emission levels by assigning vehicles the most environmentally friendly routes.

The resulting framework is capable of modeling large-scale transportation networks while capturing the inter-dependency of the transportation and communication systems. We use this framework to study the impact of communication on the feedback eco-routing application performance as an example ITS application. The following subsection describes the dynamic eco-routing logic implemented in the INTEGRATION software.

\subsection{Eco-Routing Navigation}
\label{sec:eco}
Previous studies have shown that standard navigation systems can provide travelers with accurate minimum path calculations based on either the shortest distance or the shortest travel time, and that by using these systems we can achieve some fuel savings \cite{barth2007environmentally}. However, previous studies have also shown many cases where the minimum travel time route can produce higher fuel consumption and emission levels because of the transient behavior along the route. Similarly, using the shortest travel time routes may also result in higher fuel consumption and emission levels. An example for this case was demonstrated in Ahn and Rakha \cite{ahn2008effects}, where significant savings in fuel consumption and emission levels were produced by using longer-time and shorter-distance arterial routes.

Eco-routing navigation techniques were introduced to minimize vehicle fuel consumption and emissions. Eco-routing navigation techniques utilize the route fuel cost as a metric, based on which, the most environmentally friendly route is selected. Developing and deploying eco-routing navigation techniques is very challenging. One major challenge is the estimation of the route fuel cost. This challenge comes from the fact that the route fuel cost is a function of many parameters, including the route characteristics (i.e., length, maximum speed, grade), vehicle characteristics (e.g., weight, shape, engine, and power), and driving behavior (vehicle trajectory). It has been demonstrated that it is too difficult to combine all these parameters in a single model, especially because many of these parameters are stochastic and there is a complex inter-dependency between them. 

Therefore, the best way to calculate the route cost is to use a feedback system that uses the experiences of other vehicles to compute the expected cost along a roadway segment. These data can be collected in real-time and fused with historical data to estimate the route fuel cost and consequently calculate the best route for the vehicles traveling in real-time \cite{rakha2012integration}. This feedback system is simple and accurately estimates the route cost. But, on the other hand, it requires a communication infrastructure through which these information can be exchanged. In addition to communication, vehicles should be capable of quantifying  fuel consumption for each road link.


\subsection{Eco-Routing in Literature}
Eco-routing was initially introduced in 2006 and was applied to the street network in the city of Lund, Sweden, to select the route with the lowest total fuel consumption and thus the lowest total $CO_2$ emissions \cite{ericsson2006optimizing}. The streets were divided into 22 classes based on the fuel consumption factor for peak and non-peak hours, and three vehicle classes were used. This routing technique resulted in 4\% average savings in fuel consumption. Ahn and Rakha  \cite{ahn2007field} showed the importance of route selection on the fuel and environment. They demonstrated that the emission and energy optimized traffic assignment based on speed profiles can reduce $CO_2$ emissions by 14\% to 18\%, and fuel consumption by 17\% to 25\% over the standard user equilibrium and system optimum assignment. Barth et al \cite{barth2007environmentally} attempted to minimize the vehicle fuel consumption and emission levels by proposing a new set of cost functions that include fuel consumption and the emission levels for the road links. Boriboonsomsin et al. \cite{boriboonsomsin2012eco} developed an eco-routing navigation system that uses both historical and real-time traffic information to calculate the link fuel consumption levels and then selects the fuel-optimum route. In \cite{elbery2016eco} we enhanced the eco-routing algorithm developed in \cite{ahn2007field} by introducing a new ant-colony based  updating technique for eco-routing. In 2017, we developed the first system optimum eco-routing model \cite{elbery2017novel} that uses linear programming and stochastic route assignment to minimize the system wide fuel consumption. The system optimum eco-routing reduced the fuel consumption by about 36\% compared to the user equilibrium model. 

However, all these previous efforts did not consider the communication network and its influence on eco-routing system performance. The only work that considered this impact is our previous work  in \cite{elbery2015eco} that uses discrete event simulation to model communication in VANET. However, the system developed in \cite{elbery2015eco} showed limited scalability and is not capable of modeling large-scale networks. Thus, this is the main contribution of our work in this paper. 

\subsection{Eco-routing in INTEGRATION software without Communication}
\begin{figure}[b!]
     \includegraphics[scale=0.6]{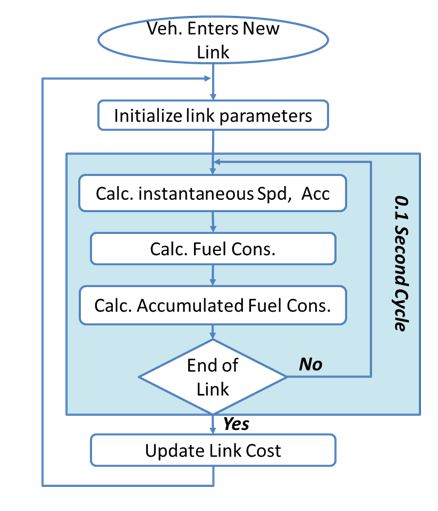}
     \centering
     \caption{Eco-routing without the Communication modeling}
     \label{fig:eco_no_com}
\end{figure}

In the INTEGRATION framework, eco-routing is developed as a feedback system that assumes that the vehicles are connected and equipped with Global Positioning Systems (GPSs) system. Moreover, vehicles are assumed to be capable of calculating the fuel consumption for each road link it traverses and communicating this information to the TMC. In INTEGRATION, the fuel consumption rate and emission rates of each vehicle are calculated every second based on the instantaneous speed and acceleration. As shown in Fig.  \ref{fig:eco_no_com}, every deci-second, the speed and the acceleration of each vehicle are calculated as well as the vehicle’s fuel consumption rate. The fuel consumption is accumulated for each road link. Then, whenever the vehicle exits that link, it updates the link cost in the TMC directly.

\subsubsection*{Estimating Vehicle Fuel Consumption}
The INTEGRATION modeler computes a number of measures every deci-second including the fuel consumed; vehicle emissions of carbon dioxide ($CO_2$), carbon monoxide (CO), hydrocarbons (HC), oxides of nitrogen ($NO_x$), and particulate matter (PM) in the case of diesel engines \cite{rakha2012integration}.

The granularity of deci-second computations permits the steady-state fuel consumption rate for each vehicle to be computed each second on the basis of its current instantaneous speed and acceleration level. INTEGRATION computes the fuel consumption and emission levels using the VT-Micro model \cite{rakha2004emission} which was developed as a statistical model from experimentation with numerous polynomial combinations of speed and acceleration levels to construct a dual-regime model, as demonstrated in Equation \ref{eq:frate}.
\begin{equation}
  \label{eq:frate}
 F(t)= \left\{\begin{matrix}
 exp\left ( \underset{i=1} {\overset{3} \sum}{ \; \underset{j=1} {\overset{3}\sum } L_{i,j}v^{i} a^{j} }\right )& if & a\geqslant 0 \\
 &  & \\
 exp\left ( \underset{i=1} {\overset{3} \sum}{ \; \underset{j=1} {\overset{3}\sum } M_{i,j}v^{i} a^{j} }\right )&  if & a<0\\
\end{matrix}\right.
\end{equation}

Where \(L_{i,j}\) are model regression coefficients at speed exponent $i$ and acceleration exponent  $j$, $M_{i,j}$ are model regression coefficients at speed exponent $i$ and acceleration exponent  $j$,   $v$ is the instantaneous vehicle speed in (km/h), and $a$ is the instantaneous vehicle acceleration (km/h/s) [39]. If route $R_n$ consists of $k$ links, the total route fuel consumption level $F_{R_n}$ is the summation of the fuel consumption of the constituting links as expressed in Equation  \ref{eq:rcost}
\begin{equation}
  \label{eq:rcost}
    F_{R_n} (t)=\sum_{m=1}^{k}F_{l_m } (t)
\end{equation}

\subsubsection*{Initial Route Assignment}
Initially, when the network is empty, the fuel consumption is computed assuming the vehicle travels at the roadway free-flow speed.  The rationale is that the routes are initially empty and the vehicles will use the roadway free-flow speed. 

This assignment will be then updated based on the information received from the individual vehicles, where each vehicle is assumed to be able to measure its fuel consumption level along each link and send this link fuel-cost to the TMC. To avoid temporal oscillations in route choices, the user can introduce a white noise error function into the link cost function. This allows vehicles to select slightly sub-optimum routes if the cost along alternative routes are very similar and thus distribute the traffic. The noise parameter also ensures that a stochastic user equilibrium is achieved.

\subsection{Eco-routing  with Wireless Communication}
\begin{figure*}
     \includegraphics[scale=0.6]{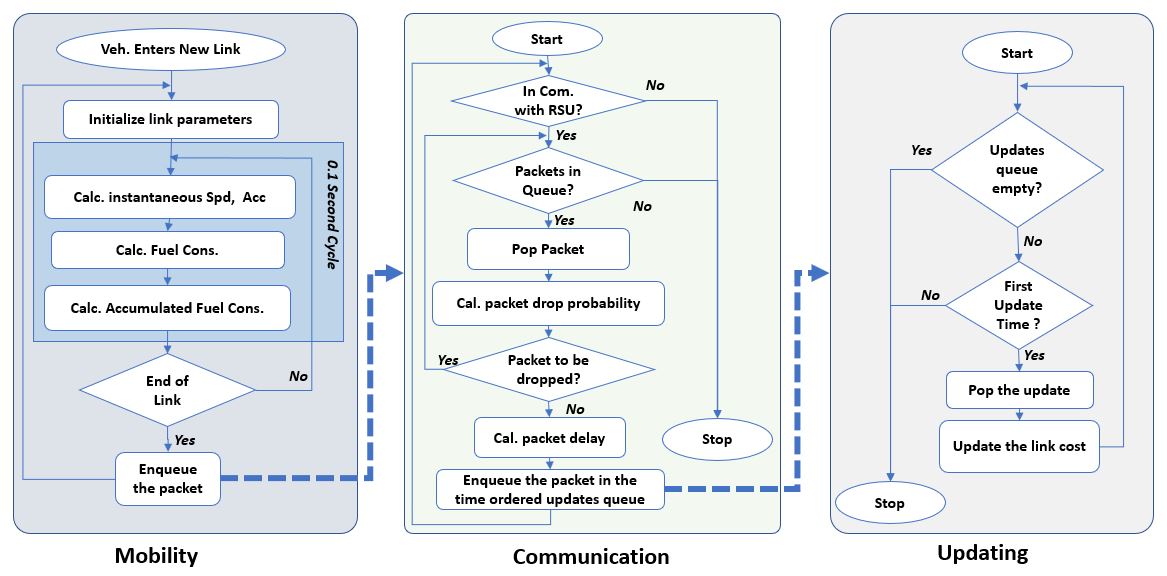}
     \centering
     \caption{Eco-routing with the communication.}
     \label{fig:eco_com}
\end{figure*}

In most of the implementations of the transportation applications, a perfect communication network is assumed, which means the drops and delays are assumed to be zero. The INTEGRATION also makes this assumption.  Thus, to build a realistic ITS application simulation framework, we modified the INTEGRATION behavior to adopt the communication performance parameters and their impact on the ITS applications. The new  behavior is illustrated in Fig. \ref{fig:eco_com}. When a vehicle finishes a road link, instead of directly updating the link cost in the TMC, the vehicle sends this information to the communication module by adding it to the transmission queue. Then, while the vehicle moves, the communication module checks for  communication network connectivity. Whenever the vehicle becomes connected to an RSU, the communication module processes the packets in the queue. For each packet, it first calculates its drop probability as described later in Equation \ref{eq:Pdrop}. If the packet should be delivered, the communication module calculates its average total delay and inserts it into a time-based ordered queue. So, it will be processed by the updating module in its time of arrival.

To calculate the packet drop probability and total delay, the communication model parameters introduced in section \ref{subsec:Derivation} are used.

With regards to the packet drop probability, the packet will be dropped in two cases. First, if it arrives at the MAC while the queue is full. The second case is when its maximum retransmission attempts is reached. If the queue is full,  the packet will be rejected by the queue. We call this the rejection probability  \(P_{rej}\). So, \(P_{rej}\) is the probability that the queue has $K$ packets. According to the $M/M/1/K$ model \cite{kleinrock1976queueing},  \(P_{rej} \) can be computed as:
 \begin{equation}
     \label{eq:Preg}
      P_{rej} =\left\{\begin{matrix}
 \rho^K \frac{1-\rho}{1-\rho^{K+1}}  &  \lambda\neq  \mu ; \\ 
  &  \\ 
 \frac{1  }{K+1}  &  \lambda =   \mu .
\end{matrix}\right.
\end{equation}

For the second case, a packet in the queue might be dropped if it experienced a collision in its last retransmission attempt whose probability is \(P(M+f-1,0) \; p_{col}\). If neither of these cases happened, then the packet will be correctly delivered. Thus, the packet drop probability can be calculated as
\begin{equation}
     \label{eq:Pdrop}
      P_{drop} = 1-(1-P_{rej})\Big(1-P(M+f-1,0) \; p_{col}\Big).
\end{equation}

The total delay of the packet \(T_{delay}\) is the summation of both service time \(T_{serv}\) and the queuing delay \(T_{q}\) as shown in Equation \ref{eq:Td}.
\begin{equation}
     \label{eq:Td}
       T_{delay} = T_{serv} + T_{q},
\end{equation}
where the average service time \(T_{serv}\) is calculated by Equation \ref{eq:Tserv}, and the average queuing delay \(T_{q}\) can be calculated  as:
\begin{equation}
     \label{eq:Tq}
      T_{q} = \frac{1}{\mu-\lambda_{eff}}
\end{equation}

In Equation \ref{eq:Tq}, \(\lambda_{eff}\) is the effective packet generation rate which is the actual number of packets that enter the queue per unit time. In other words, \(\lambda_{eff}\) is the packets that arrive when the queue is not full. If the \(Q_n\) is the probability that the queue has $n$ packets, then \(\lambda_{eff}\) can be calculated as:
\begin{equation}
     \label{eq:effrate}
      \lambda_{eff} = \lambda (1-P_{rej}) = \lambda (1-Q_K)
\end{equation}

Using Equations \ref{eq:Preg} and \ref{eq:Td} through \ref{eq:effrate}, the average total packet delay can be calculated.

\color{black}

\section{Simulation and Results}

In this paper, we used the developed model to study the mutual impacts of the communication system and the ITS feedback-based eco-routing application at different car traffic demand levels. These impacts include how the vehicle traffic congestion and flow are affected by communication errors. We also quantify the impacts  of communication errors on the fuel consumptions, the travel time, the average vehicle speed, and the  average emission levels. This section also shows how the vehicle demand level influences the communication performance in terms of packet drop rate and delay.

In this study, we use the V2I communication paradigm assuming a $1000m$ communication range and $50\; Packets/second$ background packet generation rate ($\lambda$). The average packet size is set at $1000 \; Bytes$, and the queue size is set at $64\; Packets$. In addition, we use two main communication scenarios; the ideal communication scenario assumes a perfect communication (no drops nor delay), and the realistic communication case  where the packets can be dropped and/or delayed based on the surrounding network conditions. 

The downtown area in the city of Los Angles (LA), shown in Fig. \ref{fig:RSU_Alloc}, is  used for the simulation analysis. This road network is about $133 Km^2$. It has 1625 nodes, 3561 links, and 459 traffic signals. With regards to the vehicular traffic demand, we use a calibrated traffic demand, which is based on the data collected from multiple sources, as described in detail in \cite{Janhe2018}. This traffic demand represents the morning peak hours in downtown area of the city of LA, which continues for 3 hours from 7:00 am  to 10:am. We added one hours for traffic pre-loading. So, the demand runs for four hours. However, we run the simulation for 30000 seconds to give the vehicles the enough time to finish their trips. To study the impact of different traffic origin-destination demand (OD) levels, the calibrated  traffic rates are multiplied by scaling factors (ODSFs) ranging from 0.1 through 1.0 at a 0.1 increment which generates  10 traffic demand levels. The total number vehicles that is simulated in each of these scenarios is shown in Table  \ref{tab:veh_count}.

\subsection{RSU Allocation}
Since, in this paper, we focus on the V2I communication, we have to allocate the RSUs in the network. RSU allocation is shown to be critical to the performance of the communication network and consequently the performance of the eco-routing application \cite{elbery2015eco}. The most economical method is to install the RSUs at traffic signals locations to use the already installed connections and power sources. In the road network of downtown LA, there are 459 traffic signals. Consequently, we need to identify the traffic signal locations to install the RSUs to achieve the best coverage with the minimum cost, which is a min-max coverage problem. To achieve this objective, we use a greedy algorithm shown in Algorithm \ref{alg:RSU}.

\begin{algorithm}[!b]
	\caption{Select traffic signals to install RSUs}\label{alg:RSU}
	\begin{algorithmic}[1]
		\Procedure{Select traffic signals}{} \Comment{Select the minimum number of traffic signals to install RSUs in such a way that maximizes the coverage}
		\State $S \gets \{S_i : i=1,2.... \}$
		\State $G \gets \phi$   \Comment{The initial solution}
		\While{$S\not=\phi$}\Comment{There are uncovered signals}
        \State $C_i=\{S_j \in S :D_{i,j} < R_{Com} \}$ \Comment{Recalculate the coverage}

		\State $ Select \; S_i \in S \; that \; maximizes \|C_i\|$
		\State $G\gets G \cup S_i $
		\State $S\gets S \smallsetminus C_i $
		\EndWhile
		\State \textbf{return} $G$\Comment{The selected signals}
		\EndProcedure
	\end{algorithmic}
\end{algorithm}

\begin{figure}[t!]
     \centering
	\includegraphics[width=0.45\textwidth]{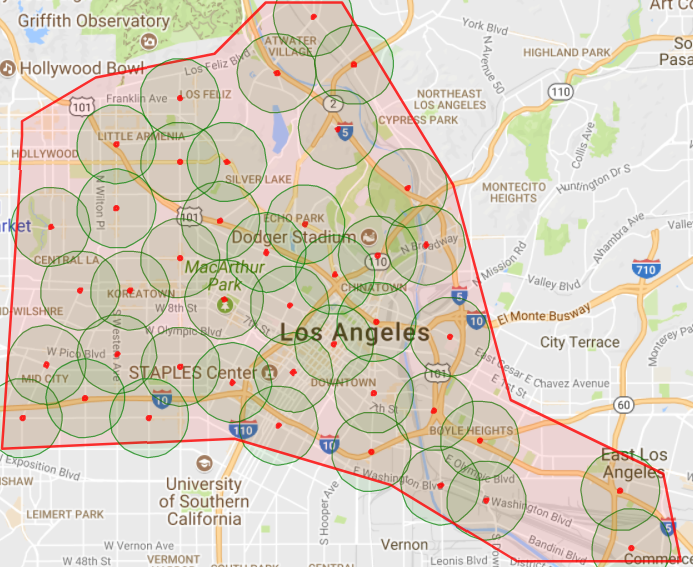}
	\caption{The LA downtown area  and the coverage map for 1000m communication ranges.}
    \label{fig:RSU_Alloc}
\end{figure}

Assuming that the distance between the traffic signals $S_i$ and  $S_j$  is $D_{i,j}$, and $C_i$ is the set of traffic signals covered by $S_i$. In other words, $C_i=\{S_j:D_{i,j} < R_{Com} \}$, where $R_{Com}$ is the communication range. The algorithm starts with $S$ includes all the traffic signals and empty set $G$ of selected signals. It calculates the coverage for each signal in $S$. Then, it selects the traffic signal that covers the maximum number of uncovered signals, add it to the selected signals $G$ and remove it along with all the signals it covers from $S$. Steps 5 to 8 are repeated until $S$ becomes empty.  This algorithm does not guarantee that the entire network will be covered, but it ensures coverage of the maximum signalized intersections with the minimum cost (minimum number of RSUs). Fig. \ref{fig:RSU_Alloc} shows the coverage map in the cases of 1000m communication range. 

\begin{figure}[!h]
	\centering
	\begin{subfigure}{0.45\textwidth}
        \centering
		\includegraphics[width=0.9\textwidth, height=1.5in]{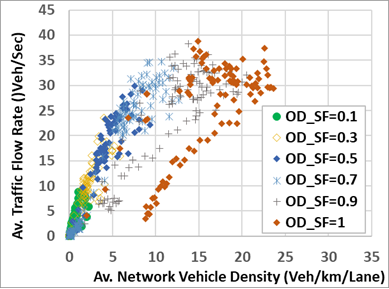}
     	\caption{}
	\end{subfigure} 
 	\begin{subfigure}{0.45\textwidth}
         \centering
		\includegraphics[width=0.9\textwidth, height=1.5in]{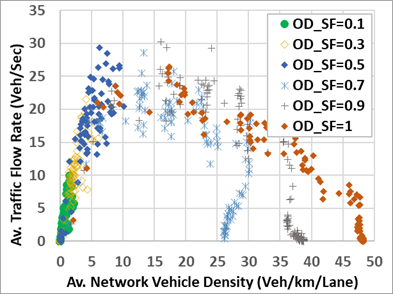}
     	\caption{}
     \end{subfigure} 
	\caption{The Network Fundamental diagrams (a) With Ideal Communication and (b) With Realistic Communication}
    \label{fig:FMD}
\end{figure}


\begin{figure*}[!ht]
	\centering
	\begin{subfigure}{0.235\textwidth}
        \centering
		\includegraphics[scale=0.455]{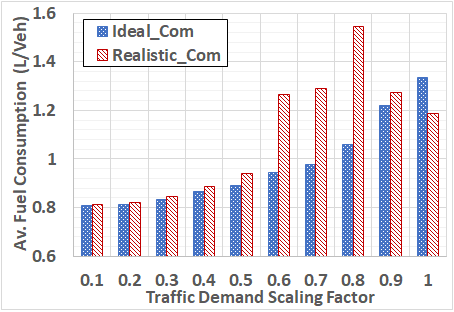}
     	\caption{}
	\end{subfigure} 
    ~
 	\begin{subfigure}{0.235\textwidth}
         \centering
		\includegraphics[scale=0.455]{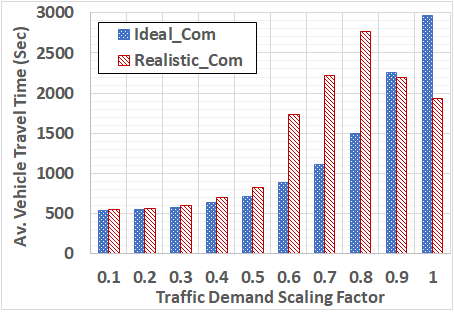}
     	\caption{}
     \end{subfigure} 
	~
	  \begin{subfigure}{0.235\textwidth}
        \centering
		\includegraphics[scale=0.455]{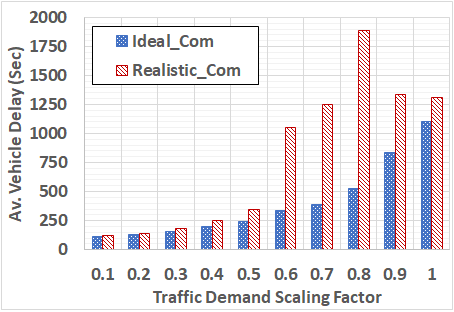}
     	\caption{}
	\end{subfigure} 
   ~
 	\begin{subfigure}{0.235\textwidth}
         \centering
		\includegraphics[scale=0.455]{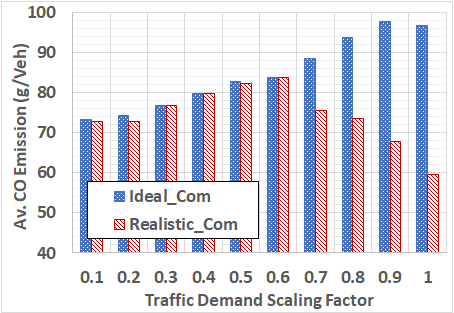}
     	\caption{}
     \end{subfigure} 
     ~
     \begin{subfigure}{0.235\textwidth}
        \centering
		\includegraphics[scale=0.455]{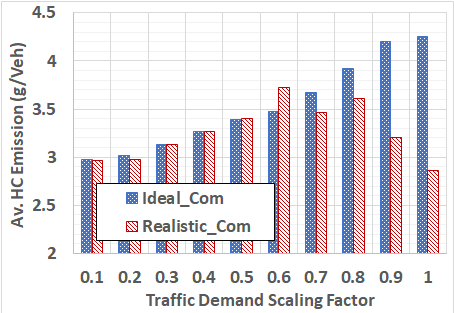}
     	\caption{}
	\end{subfigure} 
    ~
 	\begin{subfigure}{0.235\textwidth}
         \centering
		\includegraphics[scale=0.455]{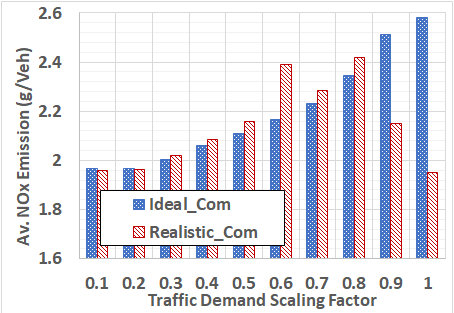}
     	\caption{}
     \end{subfigure} 
     ~
 	\begin{subfigure}{0.235\textwidth}
         \centering
		\includegraphics[scale=0.455]{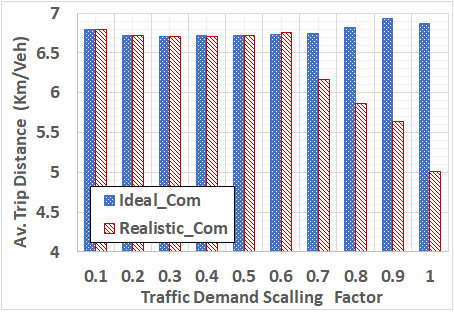}
     	\caption{}
     \end{subfigure} 
	~
	  \begin{subfigure}{0.235\textwidth}
        \centering
		\includegraphics[scale=0.455]{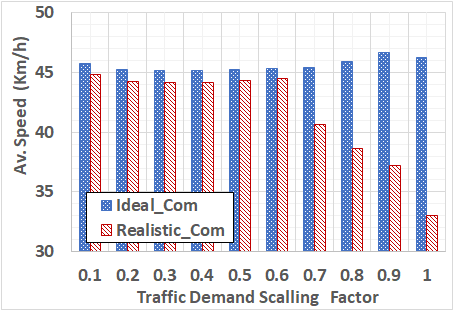}
     	\caption{}
	\end{subfigure} 
	\caption{The outputs for the ideal communication versus the realistic communication (a) The average fuel consumption per vehicle (b) the average vehicle travel time (c) The average vehicle delay (d) the average CO emission per vehicle  (e) the average HC emission per vehicle (f) the average NOx emission per vehicle (g) The average trip distance (h) the average traveling  speed}
    \label{fig:FuelDelay}
     \vspace{-4mm}
\end{figure*}

\subsection {Communication System Impact on Eco-routing System Performance}
First, we ran the network for different vehicular traffic demands levels for the case of an  ideal communication system and a realistic communication system. To identify the congestion level, we identify the network state on the Network Fundamental Diagram (NFD) \cite{helbing2009derivation} in each case as shown in Fig. \ref{fig:FMD}. The NFD shows that in the ideal communication case, the network reaches the congested regime only at an $ODSF=0.9$ and $ODSF=1.0$, while, in the realistic communication case, the congested regime takes place at a lower traffic demand level, at an $ODSF=0.7$ and higher. It also shows that there is gridlock in the network that results in a large number of vehicles being unable to exit the network at ODSFs of $0.7$ through $1.0$ in the realistic communication case. Moreover, Fig. \ref{fig:FMD} demonstrates that at the full demand level, the maximum average vehicle density is less than $25\; veh/km/lane$ in the case of the ideal communication case compared to about $47\; veh/km/lane$ in the realistic communication case, demonstrating that the congestion levels in the case of the realistic communication is approximately twice that in the case of the ideal communication scenario. 

Table \ref{tab:veh_count} shows that in the  case of ideal communication only $0.63\%$ and $2.59\%$ did not complete their trips for the two highest traffic demand levels, respectively. It also shows that $0.04\%$  and $1.02\%$ were not able to enter the network because there are no available spots to enter the network. However, in the realistic communication case at traffic scale of 0.7 about $16.88\%$ of the vehicles were not able to complete their trips, and $16.45\%$ were not able to enter the network. At the full traffic demand, it shows that  $22.41\%$ of the vehicles did not reach their final destination and $36.39\%$ had no chance to enter the network because of the congestion. This higher congestion is accompanied by lower traffic rates exiting the network in the realistic communication case. 

From Table \ref{tab:veh_count} and Fig. \ref{fig:FMD} we can conclude that the realistic communication results in packet drops and delays that lead to incorrect routing decisions that cause the network to be highly congested at a lower traffic demand compared to the ideal communication cases.

\noindent \begin{table}[h]
	\caption{Vehicles count comparison for different traffic scaling factors }	\label{tab:veh_count} 
    \centering 
	\noindent \begin{tabular}{| b | c | g |c | g |c | g |c |}
		
		\hline 
        \rowcolor{LightCyan}
		   & Total    & \multicolumn{2}{c|}{Vehicles started } & \multicolumn{2}{c|}{\% Vehicles entered  } & \multicolumn{2}{c|}{\% Vehicles   } \\  
        
        \rowcolor{LightCyan}
		OD 	& No. of &   \multicolumn{2}{c|}{ and finished } & \multicolumn{2}{c|}{\ but didn't finish  } & \multicolumn{2}{c|}{\ deferred } \\   \cline{3-8}
        \rowcolor{LightCyan}
        SF 	& vehicles &  Ideal &	Realistic &  Ideal & Realistic  &  Ideal & Realistic \\  \hline
		0.1	& 	50273 	& 	100 	& 	100    & 0 & 0 & 0 & 0\\ \hline
		0.2	& 	107047	& 	100 	& 	100   & 0 & 0 & 0 & 0\\ \hline
		0.3	& 	164499 	& 	100 	& 	100   &0  &0  & 0 & 0 \\ \hline
		0.4	& 	222326 	& 	100 	& 	100   & 0 & 0 & 0 & 0 \\ \hline
		0.5	& 	277973 	& 	100 	& 	100   & 0 &0  & 0 & 0 \\ \hline
		0.6	& 	338366 	& 	100 	& 	100   & 0  & 0 & 0 & 0 \\ \hline
    	0.7	& 	394313 	& 	100 	& 	74.67   & 0 & 16.88  & 0 & 8.45 \\ \hline
		0.8	& 	450670 	& 	100 	& 	66.87  & 0 & 21.05  & 0 & 12.08 \\ \hline
		0.9	& 	507427 	& 	99.33 	& 	60.83  & 0.63  & 20.01 & 0.04  & 19.16  \\ \hline
		1	& 	563626 	& 	96.39 	& 	41.2   & 2.59  & 22.41 & 1.02 & 36.39 \\ \hline
	\end{tabular} 
\end{table}

\subsection {Communication System Impact on Traffic Mobility}

In the realistic communication scenarios, the incorrect routing and high congestion levels resulted from the packet drops and delays are expected to result in higher fuel consumption levels, longer travel times and longer delays compared to the ideal communication case.
Fig.  \ref{fig:FuelDelay} compares the mobility parameters in the two cases for the different traffic levels.

Fig. \ref{fig:FuelDelay}-a shows that :- (1) at the low traffic demand levels, $OSDF =0.1, 0,2$ and $0.3$, the average fuel consumption is not affected by the communication system, (2) as the OSDF increases, the average vehicle fuel consumption in the case of realistic communication case becomes significantly higher than the ideal communication case, and (3) at the two highest traffic demand levels, $OSDF =0.9$ and $OSDF=1.0$, the average vehicle fuel consumption level in the realistic communication case becomes lower. The average  travel time and delay have the same behavior as shown  Fig. \ref{fig:FuelDelay}-b and Fig. \ref{fig:FuelDelay}-c. The emission levels in Fig. \ref{fig:FuelDelay}-d, Fig. \ref{fig:FuelDelay}-e, and Fig. \ref{fig:FuelDelay}-f have a similar trend.

\begin{figure}[h]
		\centering
	\begin{subfigure}{0.235\textwidth}
        \centering
		\includegraphics[scale=0.29]{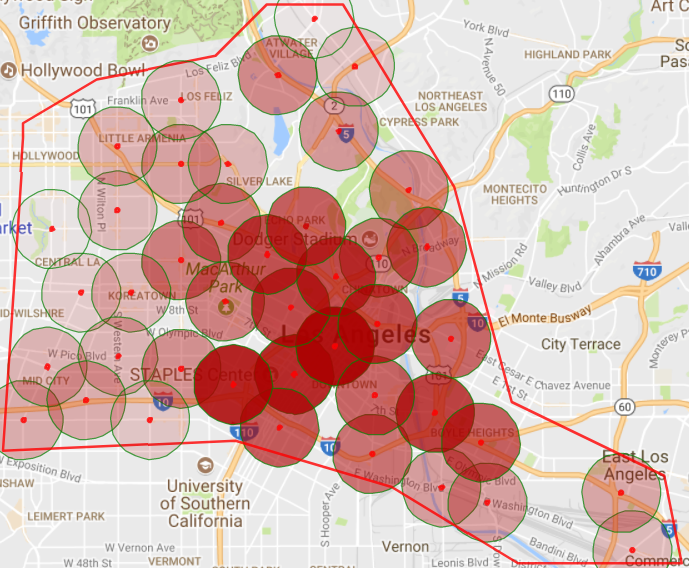}
     	\caption{}
	\end{subfigure} 
 	\begin{subfigure}{0.235\textwidth}
         \centering
		\includegraphics[scale=0.29]{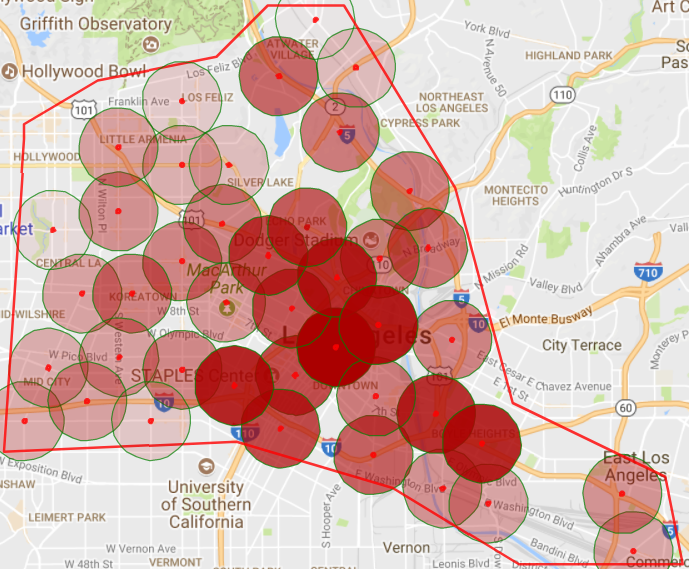}
     	\caption{}
     \end{subfigure}      
	\caption{The vehicle density at the RUS locations shown by the darkness of the red colour (a) ODSF=0.8 (b) ODSF=1}
    \label{fig:DensityMap}
\end{figure}

The third conclusion may appear counter intuitive at first glance, given that the higher vehicular traffic demand should result in higher vehicle density levels, higher packet drop rates and longer delays that should lead to incorrect routes,  higher congestion levels, and higher fuel consumption levels. These results can be attributed to two folded reasons. First, these high vehicular demand levels produce high congestion levels, as shown in Fig. \ref{fig:FMD}, consequently, a larger number of vehicles are not be able to complete their trips, as shown in Table \ref{tab:veh_count}. These vehicles are not accounted for in the fuel consumption estimates because the simulation software only includes the statistics of vehicles that complete their trips. Thus, the estimated fuel consumption counts only for $60.83\%$ and $41.2\%$ of the vehicles in the last two scenarios, respectively. The second reason is that most of the trips that finished are short trips because the short trips do not experience similar levels of congestion as do the longer trips, as shown in Fig. \ref{fig:DensityMap}. Fig. \ref{fig:FuelDelay}-g shows that the average distance per trip significantly decreased for higher traffic demand levels in the realistic communication scenarios. This shorter distance justifies the lower average vehicle fuel consumption estimates, emissions and travel times in the highest traffic demand levels.

Fig. \ref{fig:FuelDelay}-h shows the average speed in the network for the different traffic demand levels in both the ideal and the realistic communication cases. It shows that the average speed at the highest traffic level is reduced  from $46~km/h$ to bout $33~km/h$ which is also consitent with the high congestion levels in realistic communication cases.    

\subsection {Traffic Congestion Level Impact on the Communication System Performance}

The packet drop rate and delay are sensitive to the vehicle density. The higher the vehicular traffic demand, the higher the packet drop rates and the longer the vehicular delay. 

\subsubsection {Traffic Congestion Level Impact on the Communication Drop Rate}
\begin{figure}[h]
	\centering
	\begin{subfigure}{0.4\textwidth}
        \centering
		\includegraphics[scale=0.6]{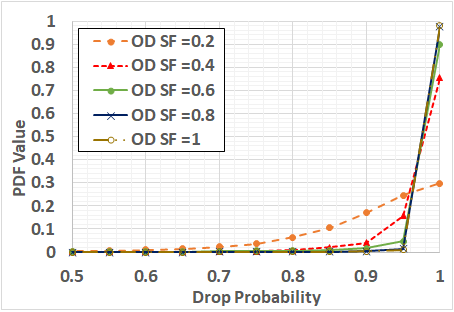}
     	\caption{}
	\end{subfigure} 
 	\begin{subfigure}{0.4\textwidth}
         \centering
		\includegraphics[scale=0.6]{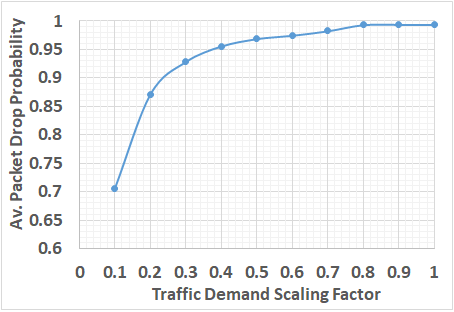}
     	\caption{}
     \end{subfigure} 
	\caption{The packet drop probability:  (a) the probability density function (pdf) of the packet drop for different traffic demand rates, (b) the average drop probability versus traffic demand rate}
    \label{fig:drop}
     \vspace{-4mm}
\end{figure}

Fig. \ref{fig:drop}-a shows the distribution of the drop probability for different traffic demand levels. It shows that the packet drop probability exponentially increases with the vehicular traffic demand level. Fig. \ref{fig:drop}-b shows the mean drop probability for each traffic demand level. Combining Fig. \ref{fig:drop} and Fig. \ref{fig:FuelDelay}, we can conclude that at an OSDF of 0.3, the eco-routing is not significantly affected by the high drop rate which reach about 93\% of the transmitted packets. 
   
\subsubsection {Traffic Congestion Level Impact on the Packet Delay}
\begin{figure}
	\centering
	\begin{subfigure}{0.4\textwidth}
        \centering
		\includegraphics[scale=0.6]{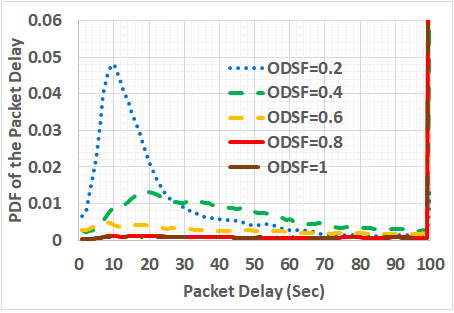}
     	\caption{}
	\end{subfigure} 
 	\begin{subfigure}{0.4\textwidth}
         \centering
		\includegraphics[scale=0.6]{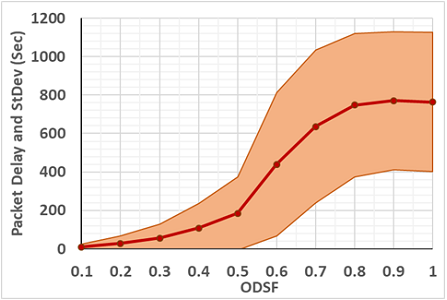}
     	\caption{}
     \end{subfigure} 
	\caption{The packet delay (Sec):  (a) the probability density function (pdf) of the packet delay, (b) the average delay and its  deviation}
    \label{fig:Packet_delay}
    \vspace{-4mm}
\end{figure}

In the vehicular environment, due to intermittent wireless connectivity, the packet queuing delay includes not only the processing delay, but also the connectivity delay (when the packet is generated when the vehicle is not connected to an RSU). This connectivity-waiting  queuing delay is dependent on the average vehicle speed, which is inversely proportional to the vehicle density. So, the higher the speed (or the lower the vehicle density) the shorter the time needed by the vehicle to reach a covered area, consequently, the  shorter the packet queuing delay.  Fig. \ref{fig:Packet_delay}-a shows the delay distribution  for different ODSF values. It shows a shorter packet delay for the low ODSF values. Fig. \ref{fig:Packet_delay}-b shows the average delay and the standard deviation area. It shows that at the high congestion levels, the mean packet delay is about $770\; seconds$ and the standard deviation is about $360\; seconds$, which means that about $4\%$ of the packets can be delayed for more than $1490 seconds$. This long delay can be attributed to the high congestion levels and the low average speed of the vehicles in these scenarios.   

\subsection {System Scalability}
\begin{figure} [!h]
     \includegraphics[scale=0.6]{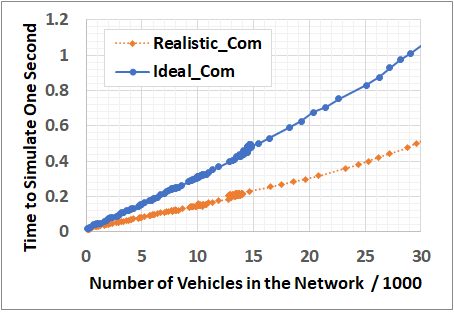}
     \centering
     \caption{Simulation speed.}
     \label{fig:sim_Speed}
     \vspace{-4mm}
\end{figure}

Since this article is interested in the large-scale modeling of VANETs, it is important to evaluate the system scalability. To achieve this objective, we calculated the time required to simulate one second for different values of concurrent vehicles on the network, which is shown in Fig. \ref{fig:sim_Speed}. The figure demonstrates that the simulation time is approximately linearly  proportional to the number of vehicles on the network. Fig. \ref{fig:sim_Speed} also shows that adding the communication modeling significantly increases the simulation time needed to simulate one second; increases the simulation time by approximately a factor of 2 when the number of vehicles in the network is $30,000$ vehicles. It is worth mentioning that this simulation completely runs sequentially, there is no parallelization utilized. So, we believe that by applying some parallel computation techniques the simulation speed can be significantly decreased.      

\section{Conclusion}
A new scalable simulation and modeling framework was developed for vehicular networks.  The proposed framework integrates  microscopic traffic modeling with a new VANET communication model that captures the mutual influence of the communication and transportation systems for the modeling of ITS applications. The proposed framework is used to quantify the impact of VANET communication on the performance of a dynamic feedback eco-routing ITS application in downtown LA using a calibrated vehicular demand. It shows that in the case of low vehicular traffic demand, the communication performance (in terms of packet drop and delay) does not have a significant impact on the dynamic eco-routing performance. However, as the vehicular traffic demand increases, this impact becomes significant. At a certain congestion level, it can result in routing gridlocks in the network due to incorrect routing decisions made by the TMC. Consequently, it is imperative to consider these mutual interactions of the communication and transportation systems when deploying such systems in real networks, especially in highly congested areas. 

The simulation results also show that the dynamic eco-routing system can work properly even at high packet drop rates that reaches approximately $93\%$ without significant performance impacts.  In other words, the dynamic eco-routing application can work properly if it receives only a small portion of the eco-routing updates correctly and in time. Consequently, it can be applied at low market penetration levels. 

In this paper, we assume V2I communication, so we did not consider  data routing within the VANET environment, which is an important component in VANETs. Consequently, in the future, we plan to study the impact of routing on the network performance. The proposed framework can also be utilized to develop novel routing protocols and study their impact on the transportation system performance for large-sale real-world scenarios.


%



\ifCLASSOPTIONcaptionsoff
  \newpage
\fi

\begin{IEEEbiography}[{\includegraphics[width=0.9in,height=1.4in,clip,keepaspectratio]{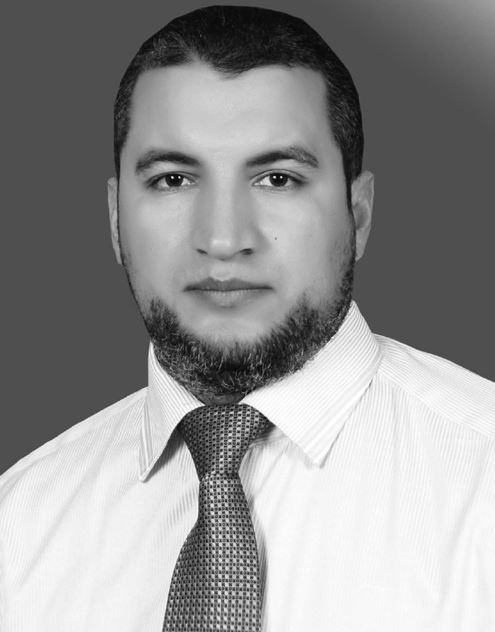}}]{Ahmed Elbery} is a postdoctoral research fellow in the Center for Sustainable Mobility (CSM) at the Virginia Tech Transportation Institute (VTTI). He received the B.Sc. degree (with honors) in Electrical and Computer Engineering from the Department of Electrical Engineering, Benah University, Egypt, and received the M.Sc. degree in Electronic Engineering from the Department of Electronic Engineering, Menoufia University, Egypt. He received his Ph.D. degree in Computer Science form the Virginia Tech, USA.  His research interests include communication and data networking, wireless and vehicular communication,  ITS, simulation and modeling of large-scale vehicular networks in smart cities, eco-routing navigation systems. He was a senior network engineer and has numerous network and communication certificates from different vendors including Cisco, Juniper, and Huawei-3Com. He also has network instructor certificates from Juniper, JNCIA, and JNCIS  Junos Enterprise Routing and Switching track. He is a Cisco Academy instructor for CCNA, CCNA wireless, and CCNA Security. 
\end{IEEEbiography}

\begin{IEEEbiography}
[{\includegraphics[width=0.9in,height=1.4in,clip,keepaspectratio]{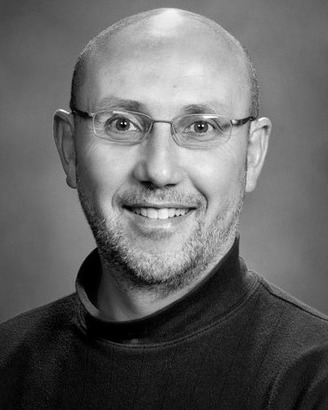}}]{Hesham Rakha}
	(M'04) received the B.Sc. degree (with honors) in Civil Engineering from Cairo University, Cairo, Egypt, in 1987 and M.Sc. and Ph.D. degrees in Civil and Environmental Engineering from Queen's University, Kingston, ON, Canada, in 1990 and 1993, respectively. He is currently the Samuel Reynolds Pritchard Professor of Engineering in the Charles E. Via, Jr. Department of Civil and Environmental Engineering, a Courtesy Professor in the Bradley Department of Electrical and Computer Engineering at Virginia Tech, and the Director of the Center for Sustainable Mobility (CSM) at the Virginia Tech Transportation Institute (VTTI). He is a Professional Engineer in Ontario and a member of the Institute of Transportation Engineers (ITE), the American Society of Civil Engineers (ASCE), the Institute of Electrical and Electronics Engineers (IEEE), the Society of Automotive Engineers (SAE), and the Transportation Research Board (TRB). He is on the Editorial Board of the Transportation Letters, IET Intelligent Transport Systems Journal, and the International Journal of Transportation Science and Technology. In addition, he is an Associate Editor for the IEEE Transactions of Intelligent Transportation Systems, the Journal of Intelligent Transportation Systems, and the Journal of Advanced Transportation. Dr. Rakha has published over 410 refereed journal and conference publications in the areas of traffic flow theory, traveler and driver behavior modeling, dynamic traffic assignment, transportation network control, use of artificial intelligence in transportation, intelligent vehicle systems, connected and automated vehicles, transportation energy and environmental modeling, and transportation safety modeling.
    \end{IEEEbiography}
    
    \begin{IEEEbiography}[{\includegraphics[width=0.9in,height=1.4in,clip,keepaspectratio]{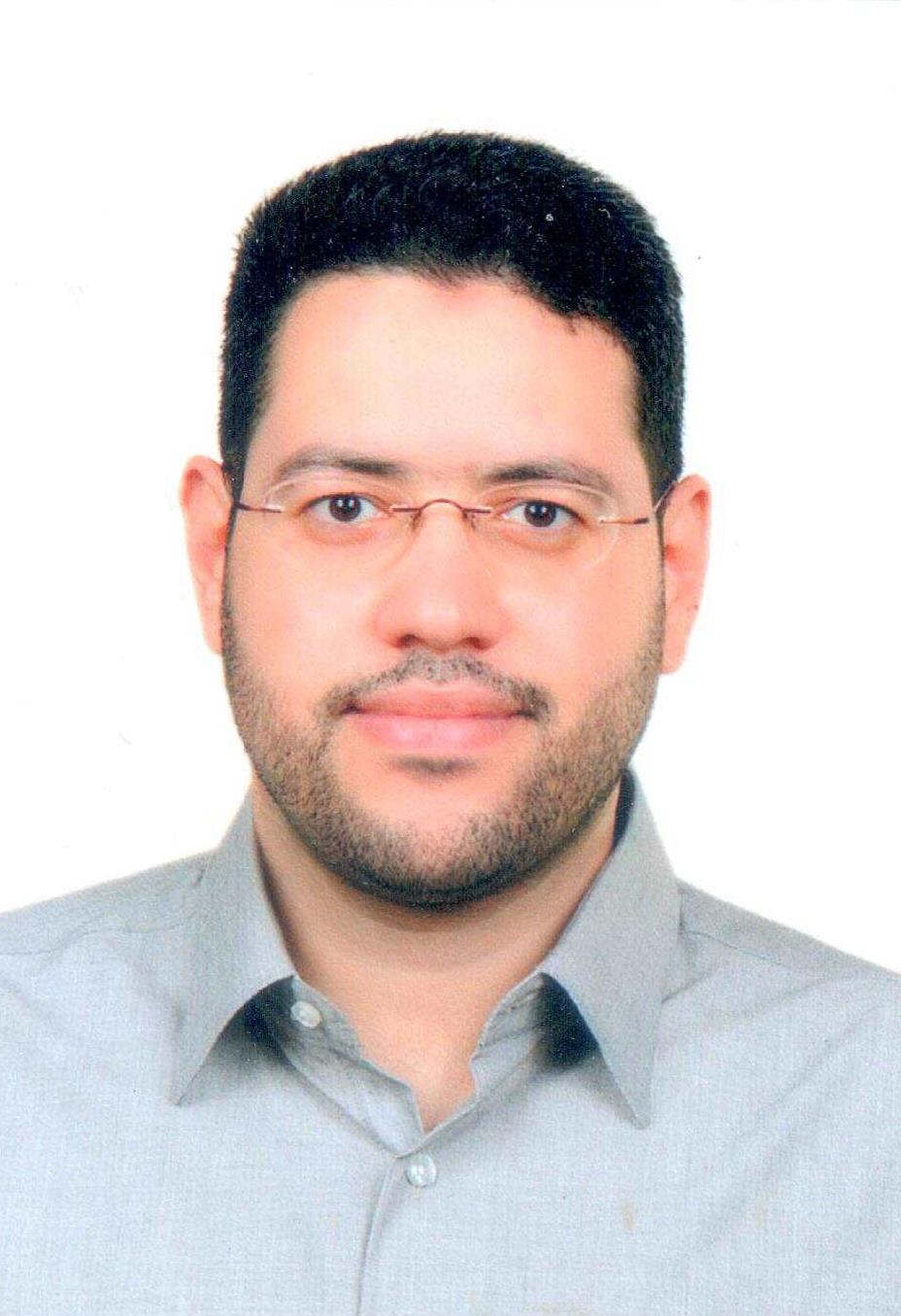}}]{Mustafa Elnainay}
    (SM'17) is an Associate Professor of Computer Engineering in the Computer and Systems Engineering department at Alexandria University, Egypt. He is also the Associate Director for administration and research of the Virginia Tech-Middle East and North Africa (VT-MENA) program and the Smart Critical Infrastructure research center (SmartCI) and Adjunct Faculty at Virginia Tech.
He received his B.Sc. and M.Sc. in Computer Engineering from Alexandria University in 2001 and 2005 respectively and his Ph.D. in Computer Engineering from Virginia Tech in 2009. His research interests include wireless and mobile networks, cognitive radio and cognitive networks, and software testing automation and optimization. Dr. ElNainay has been the PI, Co-PI, and Senior Researcher for 6 mega research projects. He is a senior member of IEEE and published more than 40 papers in international journals and conferences. He served as a reviewer and TPC member for various international journals and conferences.

%
\end{IEEEbiography}
\vfill


\end{document}